# Is the Troyon limit a beta limit?

Pierre-Alexandre Gourdain[1]†

[1]Extreme State Physics Laboratory, Department of Physics and Astronomy, University of Rochester, Rochester, NY 14627, USA

The plasma beta, ratio of kinetic to magnetic pressure, inside a tokamak should stay below the Troyon limit to avoid major plasma instabilities. However, this paper argues that Troyon limit is not strictly a beta limit, but rather an approximate limit set by calculating stability limits on equilibria which current profiles are incompatible with high beta. While the theoretical (ideal-MHD) stability of unity beta equilibrium has been resolved, the experimental existence of high beta equilibria in tokamaks remains an unsolved problem of plasma physics.

**Introduction**

Nuclear fusion embodies mankind's long lasting pursuit to a clean, mostly inexhaustible source of energy. While there can be many substitutes, nuclear fusion is a universal source of energy, the engine of the stars, the most fundamental nuclear mechanism and, of course, the most ubiquitous one. While man-made fusion reactors will use D-T in the foreseeable future, there is no reason why other fuels or non-tokamak configurations will ultimately prevail. In tokomaks, the neutron fusion power goes as (Miyamoto 1997)

$$P_n(MW) = 100\beta^2 B(T)^4 Aa(m)^3. \qquad (1)$$

$B$ is the toroidal magnetic field of the tokamak at the center of the device. $R$ and $a$ are the plasma major and minor radii, respectively. $A=R/a$ is the plasma aspect ratio. The plasma beta ($\beta$) is defined by the ratio of kinetic pressure to magnetic pressure, as given by Eq. (2).

$$\beta = \frac{2\mu_0 p}{B^2} \qquad (2)$$

This paper presents an engineering scaling of a fusion reactor to highlight that increasing the magnetic field to augment fusion throughput is not a viable solution. Then it shows that no high beta equilibrium can exist when its current profile is symmetric. As an example, the paper presents a unity beta (UB) equilibrium which goes beyond the Troyon limit, while being stable to all ideal MHD criteria, including the m=1 external kink mode. The existence of the equilibrium and its stability is simply due to the fact that the current profile is compatible with a UB equilibrium. Finally, the paper concludes that the existence of unity beta equilibrium in tokamaks is an unsolved problem of plasma physics.

**Engineering fusion scaling**

One of the largest material stresses in a tokamak can be found inside the toroidal field coil inner legs. The average centering stress $\sigma$ caused by magnetic pressure of the toroidal field on the coil inner legs is given by

†Email address for correspondence: gourdain@pas.rochester.edu

$$\sigma = \frac{B_{in}^2}{2\mu_0} = \frac{B^2}{2\mu_0}\frac{R^2}{R_{in}^2}. \qquad (3)$$

$B_{in}=BR/R_{in}$ is the toroidal magnetic field at the location (radius) $R_{in}=(R-2a)$ of the inner leg. We supposed that the thickness of the breeding blanket and vacuum vessel is on the order of the plasma minor radius $a$ here (it is ½ for ITER). Eq. (3) is a simplified version of the formula in the appendix of Duchateau *et al.* (2014), where we took $r_{ift} \sim r_{etf} \sim R_{in}$. Eq. (3) can be rewritten using the aspect ratio $A$

$$\sigma = \frac{B^2}{2\mu_0}\left(\frac{A}{A-2}\right)^2. \qquad (4)$$

We can now recast Eq. (1) using the inner leg stress from Eq. (4) instead of the magnetic field on axis.

$$P_n(MW) = 600\beta^2\sigma^2(MPa)\frac{(A-2)^4}{A^3}a^3(m). \qquad (5)$$

When the material of the toroidal field inner leg is chosen, the maximum allowable stresses are fully defined and the maximum power output of a fusion reactor is fully determined by Eq. (5). For instance, stainless-steel has a yield strength of 300 MPa. The maximum allowable stresses are 100MPa, using a safety factor of 3. This corresponds to a 16-T magnetic field on the coil. When advanced superconductors (e.g. Rare Earth, Barium-Copper-Oxide) are used, the limiting factor of a fusion reactor is its materials strength, not the critical magnetic field of the superconductor. A good example illustrating this limitation is ITER. The inner leg of the toroidal field coil has 4% of superconducting cable and 80% of stainless steel (Duchateau 2014). Therefore, a significant increase in fusion power requires an increase in: 1- plasma beta; 2- in aspect ratio $A$; 3- in minor radius $a$. The remainder of this paper discusses the high beta approach to increase fusion power.

**A necessary condition for the existence of high beta equilibrium**

According to Troyon (1984), the plasma beta is limited by its normal beta value ($\beta_N$)

$$\beta_{max} = \frac{\beta_N I}{aB} \text{ where } \beta_N \sim 2.8 \qquad (6)$$

However, the Troyon limit is not an intrinsic stability limit. Unlike the kink instability which happens under a well-defined limit (q<1), there is no hard limit for $\beta_N$. The main reason is that the Troyon limit is based on numerical simulations rather than a full, self-contained MHD stability theory. However, this "soft" limit has been observed in *all* tokamaks and its experimental validation made the Troyon limit an acceptable figure of merit to evaluate tokamak performances. However, this paper demonstrates that large $\beta$ can be stable, independent of $\beta_N$. In fact, it is well-known that the Troyon limit has changed over the years, by improving experimental techniques which increased the $\beta_N$, as shown for instance by Ferron et al. (2005). But the exploration of plasma stability was done with equilibria incompatible with large betas.

To find equilibria able to sustain high betas, we need to recast the Grad-Shafranov equation. We use the current density on the high field side of the plasma and the low field side of the plasma, $J_H$ and $J_L$ respectively, to characterize high beta equilibria. Without loss of generality, we take both current densities to be positive. As shown in Gourdain (2007), we can write the plasma pressure $p$ and the toroidal function $F$ as

$$\frac{dp}{d\psi} = \frac{R_L J_L - R_H J_H}{R_L^2 - R_H^2} \tag{7}$$

and

$$\frac{dF^2}{d\psi} = 2\mu_0 R_L R_H \frac{R_L J_H - R_H J_L}{R_L^2 - R_H^2} \tag{8}$$

$R_H$ is the major radius on the high field side of the flux surface with flux $\psi$ where $J_H$ flows and $R_L$ is the major radius on the low field side of the same flux surface where $J_L$ flows, both on the plasma mid-plane (which contain the magnetic axis). It is always possible to rewrite $dp/d\psi$ and $dF^2/d\psi$ in this manner.

Both equations are axisymmetric equilibrium equations. Any departure from these equations and the equilibrium will cease to exist. A necessary and sufficient condition to reach high beta is to increase pressure gradient $dp/d\psi$ (i.e. increase $p$ or decrease $\psi$). As Eq. (7) shows there are only two options to increase $dp/d\psi$ at equilibrium in a finite aspect ratio tokamak (i.e. $R_H < R_L$). The first option is to reduce $R_H$ as much as possible. This is the option chosen by spherical tokamaks like NSTX or MAST. However this path is not a valid approach to fusion reactors since it brings the aspect ratio $A$ close to 2. As Eq. (5) shows, the fusion power of a spherical tokamak would be drastically low. The second option is a strongly asymmetric current density profile (i.e. $J_L >> J_H$).

As a result, increasing the plasma beta without making the current profile strongly asymmetric leads to the non-existence of an equilibrium rather than an equilibrium instability. In his seminal paper, Troyon studied the plasma beta of tokamak equilibria with quasi-symmetric current density profiles. Since these current profiles can only support low beta equilibria, Troyon *et al.* (1984) could not have found any stable high beta equilibria. And they did not, hence the Troyon limit.

**An example of stable unity beta equilibrium**

On the other hand if we allow strong asymmetry in the current profile, it is possible to generate UB equilibria. FIGURE **1** shows an example of UB equilibrium using ITER's geometry. The peak $\beta$ (i.e. on axis) is unity. The average $\beta$ is 13% and the $\beta_N$ is 5.5. DCON (Glasser 1995) showed the stability of this equilibrium to internal *and* external kink modes from n=1 to 4, as well as high-n ballooning and GGJ resistive modes.

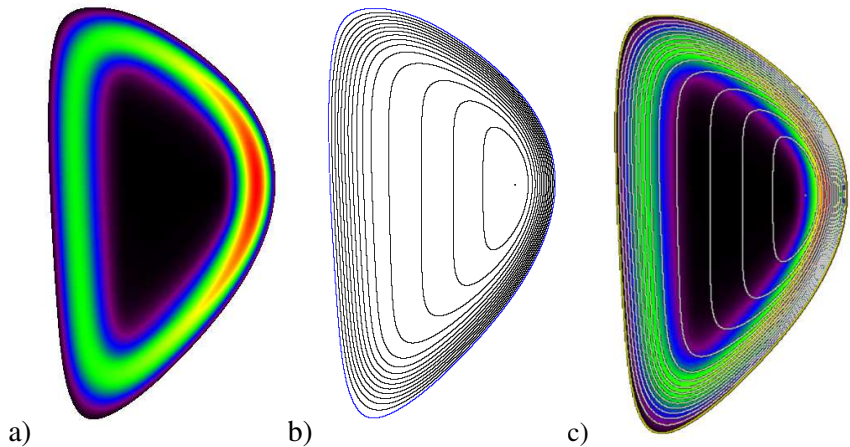

a) b) c)

FIGURE 1. Unity beta tokamak equilibrium using ITER's geometrical parameters and toroidal field. a) toroidal current profile, b) flux surface distribution and c) a composite image of both toroidal current profile and flux surface distribution.

The plasma beta is increased in the equilibrium of FIGURE **1** simply by setting $J_H$=0 while keeping $J_L$ large in the central plasma region, encompassing the five most inner flux surfaces. The outer region of the plasma, where $J_H \sim J_L$ is just there to stabilize external kink modes by blanketing the high beta core with a paramagnetic plasma, which restore the external kink stability. To this extend, the outer plasma blanket acts as a perfectly conducting wall to the high beta core, stabilizing it. While the current profile shown in FIGURE **2** is reminiscent of the current hole (CH) equilibrium found in JET (Hawkes 2001) and JT-60 (Fujita 2001), the equilibrium presented herein is not a CH equilibrium.

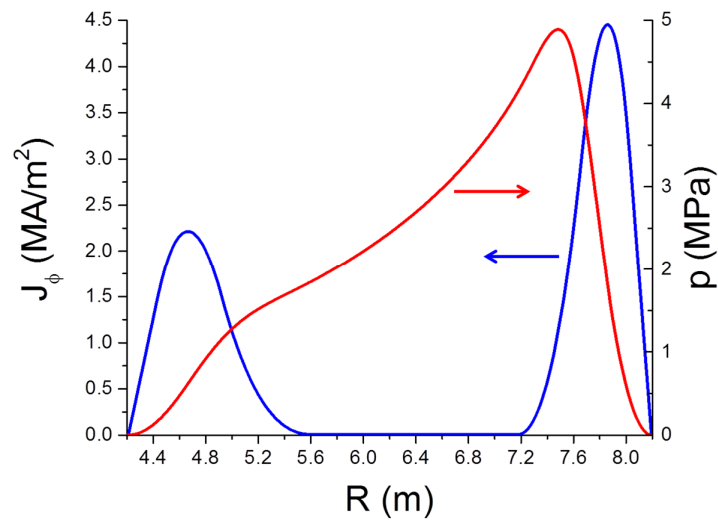

FIGURE 2. Current ($J_\phi$) and pressure (p) profile for the equilibrium shown in FIGURE 1.

CH equilibria are paramagnetic and have no magnetic flux surfaces inside the region where the plasma current density is zero. FIGURE **1** shows magnetic flux surfaces inside the region where toroidal current density is zero. Unlike in CH equilibria, diamagnetic poloidal currents flow in this region. Further there is no pressure gradient inside the hole of the CH equilibrium. FIGURE 2 clearly shows that there is a pressure gradient and this gradient is key in increasing the plasma beta.

Due to the similarity of their toroidal current profiles, it seems possible to turn a CH into a UB equilibrium with limited current drive. Starting with a CH at low toroidal field (say 10% of the nominal toroidal field of the device), one could reach a UB equilibrium by ramping the toroidal field up. This would generate diamagnetic poloidal currents inside the plasma. The low-field side toroidal current density should be increased using neutral beam driven currents. The plasma temperature and density would have to be raised accordingly using neutral beams and auxiliary heating.

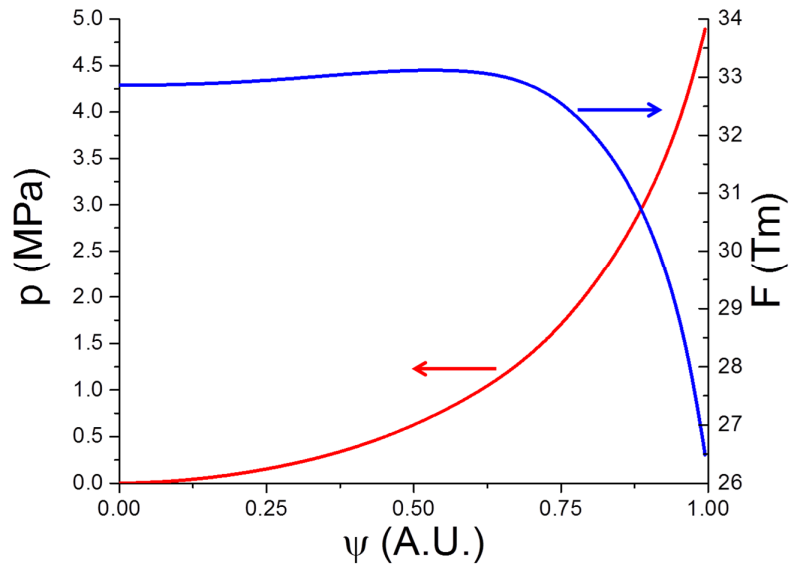

FIGURE 3. Plasma pressure (p) and toroidal function (F) as a function of the normalized flux $\psi$

The equilibrium of FIGURE 1 was obtained using CUBE (Code for Unity Beta Equilibria) which uses a multi-grid implementation to solve the Grad-Shafranov equation with free boundary. The details of the convergence at high beta are given in Gourdain (2006). The current profile and shape of the last-closed flux surface were used as inputs to the code. FIGURE 3 shows the pressure profile and toroidal function corresponding to the UB of FIGURE 1. The plasma current used was 25 MA and the toroidal field was 5.3 T at the center of the machine. The free boundary shape was obtained by using a major radius of 6.2 m and a minor radius of 2 m. The elongation factor is 1.8, the triangularity is 0.6 and the up-down asymmetry factor is -0.23.

It is important to note that tokamak plasmas with a no-wall $\beta_N > 4$ and quasi-symmetric current profiles have been studied experimentally (Taylor 1991). This is another example which demonstrates that the Troyon limit is not directly related to a beta limit. Larger $\beta_{NS}$ have been found experimentally by stabilizing the external kink mode using rotation (Garofalo 2002) or active feedback.

**Conclusion**

This paper showed that the Troyon limit arises when one tries to increase the pressure inside a plasma where the current profile cannot sustain high beta equilibrium. Using only the insight given by Eq. (7), we have shown in FIGURE **1** an example of stable equilibrium which has no beta limit and yet goes beyond the Troyon (no wall) limit. Even higher betas can be reached without encountering any stability limits. Since material strength limits the toroidal magnetic field a reactor can sustain (and to a lesser extend its size), increasing the plasma beta is another route to increase drastically fusion power for a given device and, to this day, has never been explored in conventional tokamaks, since the evaluation of high beta stability in tokamaks can only be done with highly asymmetric current profiles. Until this is achieved experimentally, the stability of high beta plasmas will remain an unsolved problem of plasma physics.